\documentclass[aps,prl, twocolumn, amsmath, amssymb, floatfix, showpacs]{revtex4-1}
\usepackage[final]{graphicx}
\usepackage{hyperref}
\usepackage{subfigure}
\usepackage{enumerate}
\graphicspath{{Figures/}}

\begin{document}
\title{Effects of Strain and Buffer Layer on Interfacial Magnetization in Sr$_2$CrReO$_6$ Films Determined by Polarized Neutron Reflectometry}
 \author{Yaohua~Liu$^{1}$}  \email[]{yhliu@anl.gov}
\author{J.~M.~Lucy$^{2}$} \author{A.~Glavic$^{3}$}\author{H.~Ambaye$^{3}$}  \author{V.~Lauter$^{3}$}\author{F.~Y.~Yang$^{2}$} \author{S.~G.~E.~te~Velthuis$^{1}$}

\affiliation{$^{1}$Materials Science Division, Argonne National Laboratory, Argonne, Illinois 60439, USA}
\affiliation{$^{2}$Department of Physics, The Ohio State University, 191 W. Woodruff Ave., Columbus, Ohio 43210, USA}
\affiliation{$^{3}$Quantum Condensed Matter Division, Oak Ridge National Laboratory, Oak Ridge, TN 37831, USA}

\date{\today}

\begin{abstract}
We have determined the depth-resolved magnetization structures of a series of highly ordered Sr$_{2}$CrReO$_{6}$ (SCRO) ferrimagnetic epitaxial films via combined studies of x-ray reflectometry, polarized neutron reflectometry and SQUID magnetometry. The SCRO films deposited directly on (LaAlO$_3$)$_{0.3}$(Sr$_2$AlTaO$_6$)$_{0.7}$ or SrTiO$_{3}$ substrates show reduced magnetization of similar width near the interfaces with the substrates, despite having different degrees of strain. When the SCRO film is deposited on a Sr$_{2}$CrNbO$_{6}$ (SCNO) double perovskite buffer layer, the width of the interfacial region with reduced magnetization is reduced, agreeing with an improved Cr/Re ordering.  However, the relative reduction of the magnetization averaged over the interfacial regions are comparable among the three samples.  Interestingly, we found that the magnetization suppression region is wider than the Cr/Re antisite disorder region at the interface between SCRO and SCNO.  
\end{abstract}
\maketitle

The family of the A$_2$BB$'$O$_6$ double perovskites includes several functional magnetic materials with both a Curie temperature (T$_C$) much higher than room temperature and a high spin polarization of the charge carriers~\cite{serrate2007double}. Ferrimagnetic Sr$_2$CrReO$_6$ (SCRO), with a T$_C$ above 500~K~\cite{kato2002metallic, de2005investigation}, is of particular interest because of its intrinsic semiconducting nature at room temperature in highly ordered epitaxial films~\cite{hauser2012fully}.  There is a strong interest in spin injection into semiconductors~\cite{jeon2014voltage, van2012low, raman2011new, barraud2010unravelling} since Datta and Das proposed a spin transistor based on electric-field controlled spin precession via spin-orbit coupling~\cite{datta1990electronic, rashba1960properties}.  However, spin injection directly from ferromagnetic metals into semiconductors is problematic due to the conductivity mismatch problem~\cite{schmidt2000fundamental}. One solution is to use magnetic semiconductors rather than metals~\cite{smith2001electrical}, thus Sr$_2$CrReO$_6$ is clearly a promising candidate.  Furthermore, the magnetic properties of SCRO are in principle very responsive to lattice changes because it is a highly correlated material with a strong spin-orbit coupling from the Re 5$d$ orbitals. Thus, SCRO can be potentially used for low-dissipation magnetoelectric devices by utilizing strain-mediated magnetoelastic coupling~\cite{komelj2010magnetoelasticity, czeschka2009giant}.

The electronic and magnetic properties of double perovskites strongly depend on the B/B$'$ ordering~\cite{serrate2007double}. For example, Cr/Re antisite disorder reduces the high spin-polarization of Sr$_2$CrReO$_6$~\cite{kato2002metallic}.  However, fabrication of high-quality double perovskite epitaxial films with full B/B$'$ ordering is still challenging. We have recently achieved phase pure SCRO epitaxial films with a high degree of Cr/Re ordering on SrTiO$_3$ (STO) substrates by a new sputtering technique~\cite{hauser2012fully, hauser2013electronic}. High-angle annular dark-field scanning transmission electron microscopy (STEM) studies show that the full Cr/Re ordering takes a few nanometers to establish when the films are directly grown on STO~\cite{lucy2013buffer}, although SCRO and STO have well-matched lattice constants. Previous STEM studies did not quantify the degree of the Cr/Re antisite disorder or how much the interfacial magnetization is affected. These are in fact very important for applications because the performance of spintronics devices crucially depends on the interfacial properties~\cite{tsymbal2007interface, barraud2010unravelling, liu2009correlation, liu2013emergent}. To this end, we have utilized x-ray reflectometry (XRR) and polarized neutron reflectometry (PNR) to study a series of SCRO films to investigate how antisite disorder, as well as strain, affects the interfacial magnetization.  We found that the  films show a 5-6 nm interfacial layer with reduced magnetization when grown directly on (LaAlO$_3$)$_{0.3}$(Sr$_2$AlTaO$_6$)$_{0.7}$ (LSAT) and SrTiO$_{3}$ substrates.  For the film grown on a SrCr$_{0.5}$Nb$_{0.5}$O$_{3}$ (SCNO) double perovskite buffer layer on STO, the width of the region with reduced magnetization decreases to 3.7~nm, which is however wider than the Cr/Re antisite disorder region, which is about 1.5~nm, at the interface between SCRO and SCNO, as found in previous work~\cite{lucy2013buffer}. On average, the relative decreases of the magnetization in the interfacial regions are comparable among the three films. 

\begin{figure}[t]
	\centering
		\includegraphics[width=0.44\textwidth]{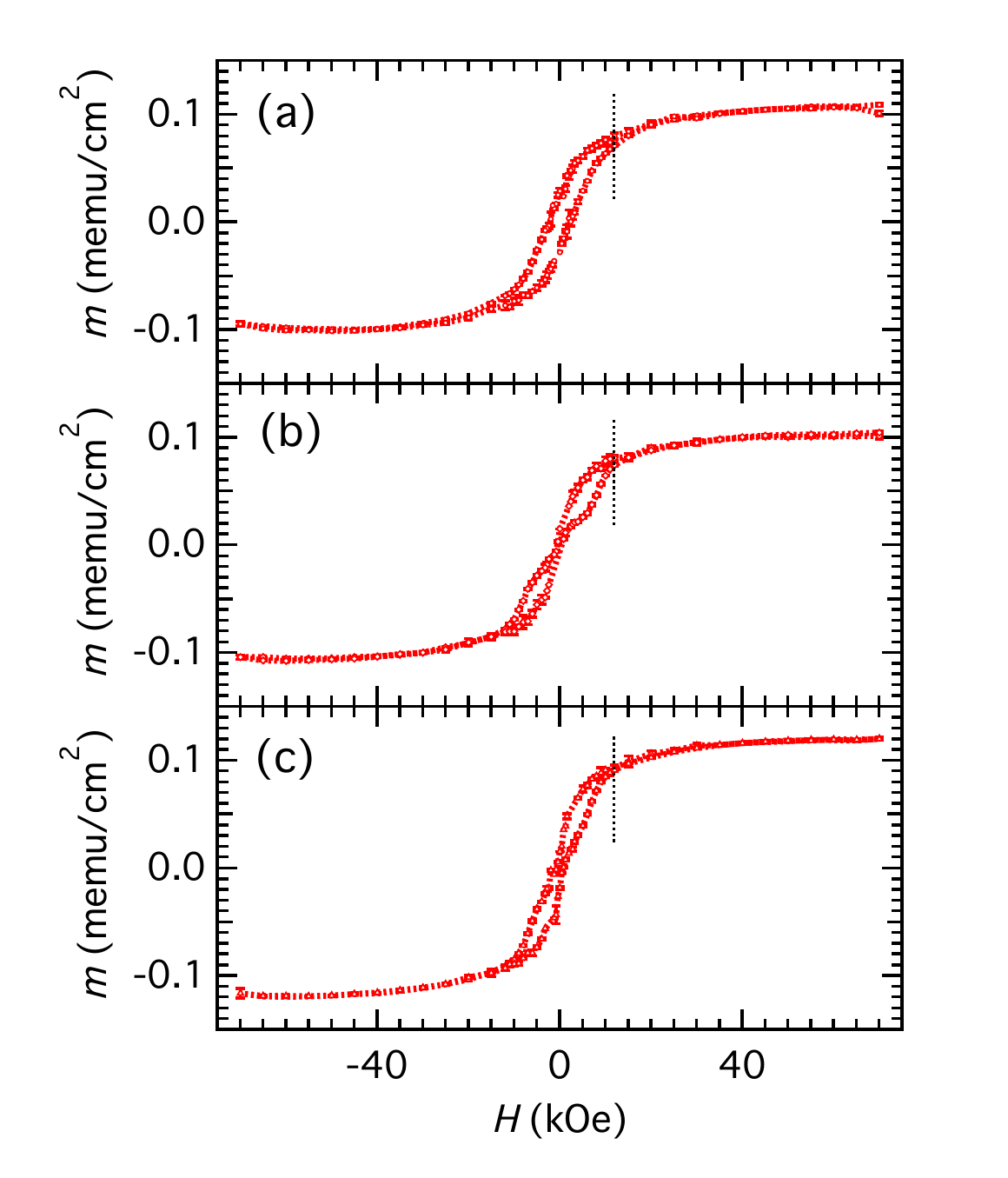}
	\caption{\label{Fig:SQUID}(Color online) Magnetic hysteresis loops from the three SCRO films grown on (a) LSAT, (b) STO and (c) a SCNO buffer layer, respectively. Data were collected between $+$70 and $-$70~kOe at 300~K, and have been normalized by the area of the films. PNR experiments were performed at 11.5~kOe, as indicated by the dashed lines.} 
\end{figure}

\begin{figure}[t]
	\centering
		\includegraphics[width=0.46\textwidth]{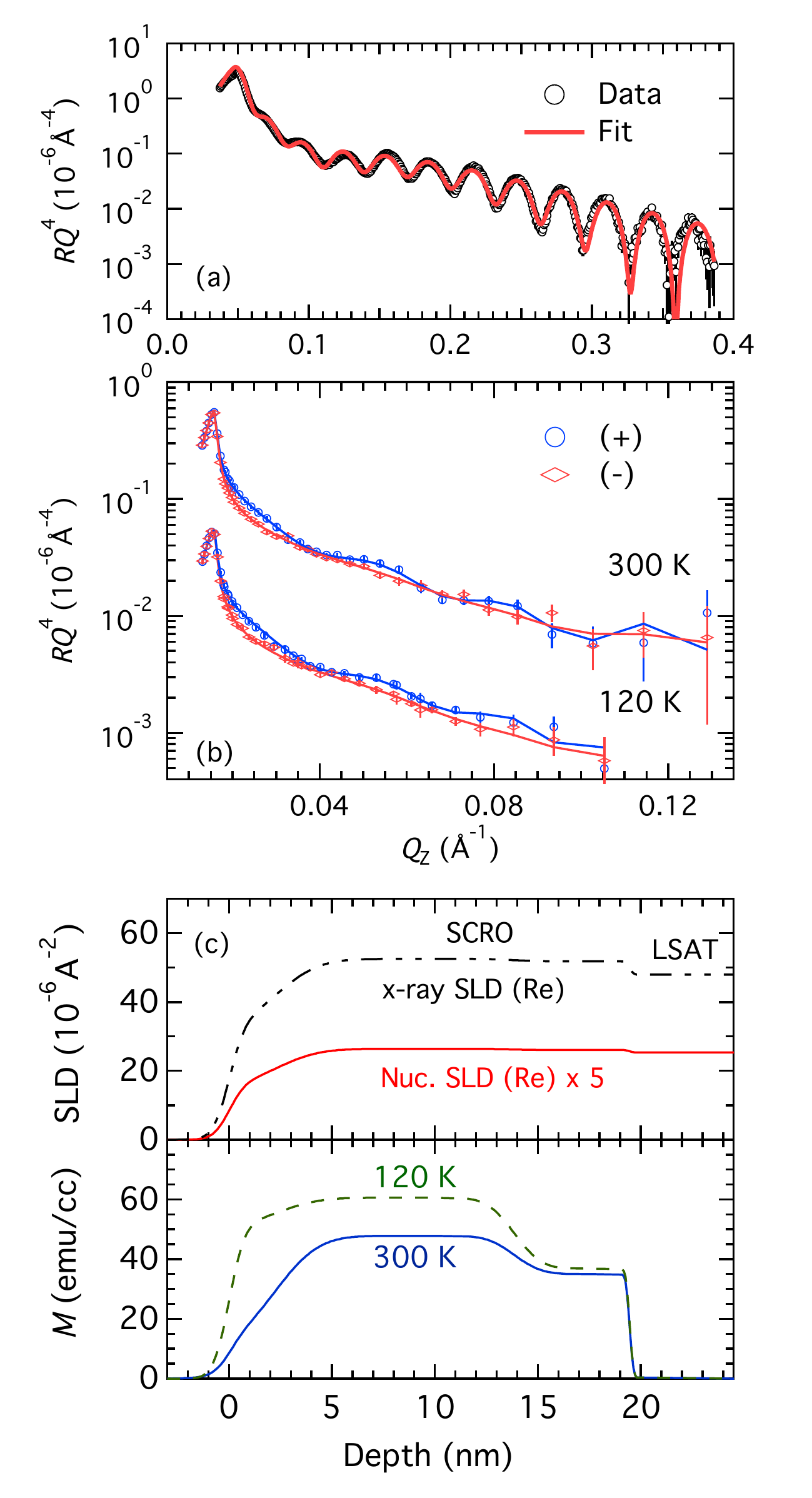}
	\caption{\label{Fig:LSAT}(Color online) Results on the 20~nm SCRO  film on LSAT. (a) X-ray reflectivity, and (b) polarized neutron reflectivity at 300~K and 120~K in a 11.5~kOe in-plane magnetic field. The best fit curves (lines) are overlaid on the top of the data (circles). (c) The depth profiles of the real parts of the x-ray and nuclear scattering length densities (SLDs, top) and the magnetization ($M$, bottom).} 
\end{figure}

A series of 20-nm (001) Sr$_2$CrReO$_6$ epitaxial thin films were grown on LSAT and STO substrates, and on a 50-nm nonmagnetic fully strained SCNO buffer layer on STO via an off-axis ultrahigh vacuum sputtering technique~\cite{hauser2012fully, lucy2013buffer}. The substrates were cleaned via sonication in acetone before being loaded into the sputtering chamber. The substrates were further baked at the deposition temperature for about 30 minutes before film deposition. A DC power source was used for the SCRO film deposition, during which the substrate temperature was 700~$^\circ$C and the total O$_2$/Ar gas pressure was 12.5~mTorr with an oxygen partial pressure of 26.25~$\mu$Torr. A radio-frequency  power source was used for the SCNO buffer layer deposition, during which the substrate temperature was 600~$^\circ$C and the total O$_2$/Ar gas pressure was 12.5~mTorr with an oxygen partial pressure of 15.5~$\mu$Torr. The deposition rates were 1.10 and 1.25~nm/min for SCRO and SCNO films, respectively. X-ray diffraction studies show that all three 20-nm films are fully strained~\cite{lucy2013buffer}. The SCRO films on STO and on SCNO are nominally strain free, while SCRO on LSAT is under 1.5\% in-plane compressive strain.  

Figure~\ref{Fig:SQUID} shows the magnetic hysteresis loops for the three samples in the field range of $\pm$ 70~kOe at 300~K.  A linear diamagnetic background has been subtracted for each sample, using the slope obtained by fitting the high-field region of the hysteresis loop.  At 11.5~kOe, which is the maximum field achievable during the PNR experiments, the loops are essentially closed at 300~K and are open at 120~K (data not shown) for all three samples. 

The three films have been further characterized via a complementary approach of x-ray reflectometry and polarized neutron reflectometry.  We performed PNR experiments at the Spallation Neutron Source at Oak Ridge National Laboratory. XRR data were collected with Cu $K_{\alpha}$ radiation at room temperature. PNR is capable of resolving the depth profile of magnetization at nanometer resolution~\cite{FelcherRSI1987,  MajkrzakPhysB1996}.  Neutrons interact with both nuclei and the internal magnetic field. Neutrons with spin parallel (Spin-up (+)) or anti-parallel (spin-down (-) ) to the direction of the external magnetic field experience the same nuclear scattering potential, but opposite magnetic scattering potentials.  From subsequent measurements with oppositely polarized neutron beams, the two contributions can be separated to reconstruct the depth profiles of both the chemical structure and the magnetization vector. The magnetic scattering length density is related to the magnetization by a constant $C = 2.91 \times 10^{-9}$ \AA$^{-2}$G$^{-1}$.  PNR data were collected with a 11.5~kOe in-plane magnetic field at 300~K, and at 120~K after field cooling. The magnetic field was applied along one of the sample edges ([100] direction). A polarization analyzer was not used during the experiments. The selected temperature of 120~K is well below the T$_C$ of these films, and high enough so that we can avoid the loss of specular reflectivity due to the surface buckling after the cubic to tetragonal phase transition of the SrTiO$_3$ substrate near 105~K~\cite{cowley1969relationship}. XRR and PNR data were fitted with the same model for the chemical structural, and the Parratt formalism was used to determine the depth-dependent scattering length density (SLD) profiles~\cite{ParrattPR1954, liu2011magnetic}. Levenberg-Marquardt least-square method implemented by IGOR Pro was used for model optimization~\cite{IGORPro}.  Error bars for the data points and the structural profiles are determined by the statistical uncertainties.

The XRR and PNR data from the film on LSAT are shown in Figs.~\ref{Fig:LSAT}(a) and~\ref{Fig:LSAT}(b), respectively.  Simulated reflectivity curves from the best fit are displayed as solid lines overlaid on the top of the data. The polarized neutron reflectivity spans about 6~orders of magnitude and the x-ray reflectivity spans about 7~orders of magnitude. Due to the large dynamic ranges, the reflectivity data were presented as $RQ^{4}$ (logarithmic scale) vs $Q$ for better visualization. $R$ is the specular reflectivity and  $Q = 4 \pi sin\theta /\lambda$, where $\theta$ is the incident angle and $\lambda$ is the wavelength of the radiation source. XRR data show pronounced oscillations due to the high contrast in the x-ray scattering length densities between SCRO and LSAT.  The oscillation period is determined by the total film thickness, which is 19.5~nm. For neutrons SCRO and LSAT have a weak contrast in nuclear scattering length densities. Thus, the oscillations are largely caused by the contrast in magnetization, which is advantageous for determining the details of the film's magnetization structure.  In the model, the SCRO film is split into three layers (two interfacial regions and the central part), each of which has its own scattering length density and magnetization, as well as thickness.  A rough interface was modeled as a sequence of thin slices with SLDs values varying as an error function so as to interpolate between adjacent layers~\cite{liu2011magnetic}.  The XRR plays a more important role in determining the chemical structure of the film. This is because that the XRR data were collected up to a larger $Q$ and with a better statistics than the PNR data, and the contrast in the x-ray SLDs between SCRO and LSAT is higher than that in the neutron SLDs. The algorithm used for PNR simulations assumes that the average magnetization component perpendicular to the field is zero.  Based on the hysteresis loops,  this approximation is more accurate at 300~K than at 120~K; thus, the fitting results from the 300~K data are more reliable.

Figure~\ref{Fig:LSAT}(c) shows the depth profiles of the neutron nuclear scattering length density as well as the magnetization at 120~K and 300~K from the best fit.  In order to describe the XRR data,  it is necessary to include a surface layer of 2.3~nm in the model, which has a lower x-ray scatting length density than the bulk part of the SCRO film. Note that surface layers have been observed in related double-perovskite Sr$_2$FeMoO$_6$ films after being exposed to air~\cite{rutkowski2010x, jalili2009x}. More importantly here, the SCRO film shows a region of 5.6~nm with reduced magnetization at the interface with STO.  The average magnetization of the interfacial region is $73 \pm 6 \%$ and $60 \pm 9 \%$ of the magnetization of the bulk part of the film at 300~K and 120~K, respectively.  When temperature decreases from 300~K to 120~K, the magnetization of the whole film increases by 30\%. A similar magnitude change is also observed from the SQUID magnetometery data. However, a strict comparison is not meaningful because the hystersis loop is not closed in a 11.5~kOe field at 120~K.  

\begin{figure}[t]
	\centering
		\includegraphics[width=0.46\textwidth]{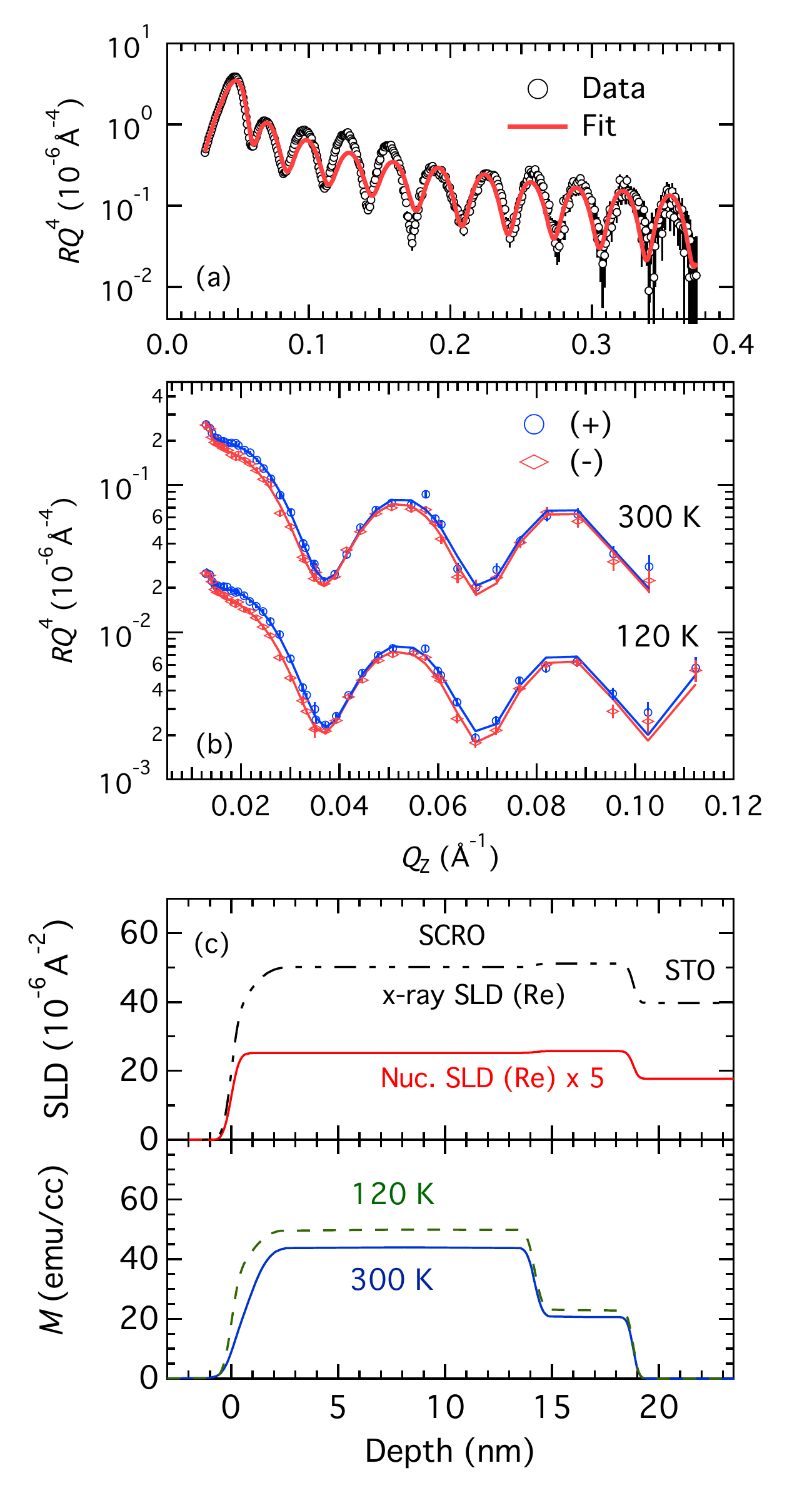}
	\caption{\label{Fig:STO}(Color online) Results on the 20~nm SCRO  film on STO. (a) X-ray reflectivity, and (b) polarized neutron reflectivity at 300~K and 120~K in a 11.5~kOe in-plane magnetic field. (c) The depth profiles of the real parts of the x-ray and nuclear scattering length densities (top) and the magnetization (bottom).} 
\end{figure}

\begin{figure}[t]
	\centering
		\includegraphics[width=0.448\textwidth]{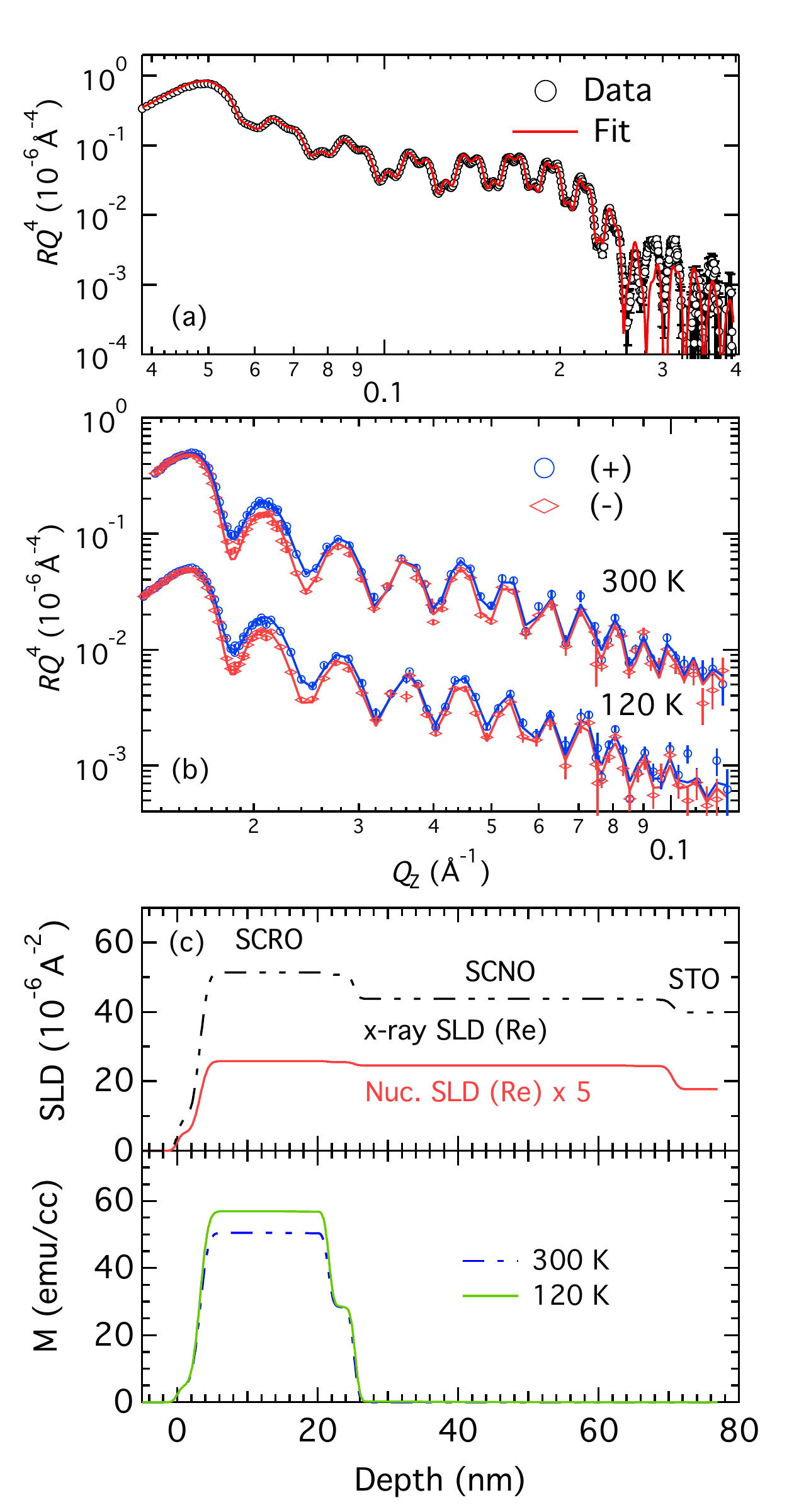}
	\caption{\label{Fig:STObuff}(Color online) Results on the 20~nm SCRO  film on a 50-nm SCNO buffer layer on STO. (a) X-ray reflectivity, and (b) polarized neutron reflectivity at 300~K and 120~K in a 11.5~kOe in-plane magnetic field. (c) The depth profiles of the real parts of the x-ray and nuclear scattering length densities (top) and the magnetization (bottom).} 
\end{figure}

The results from the SCRO film on STO are shown in Fig.~\ref{Fig:STO}. It is obvious that although both SCRO films on LSAT and STO are nominally of 20~nm thickness, the reflectivity curves (both XRR and PNR) from the two samples are dramatically different. This difference reflects the fact that reflectivity is highly sensitive to the depth profile of the scattering length density.  In contrast to the SCRO film on LSAT, the oscillation magnitude of the PNR data on the SCRO film on STO is very large, which is mostly due to the high contrast in the nuclear scattering length densities between the SCRO and STO. As shown in Fig.~\ref{Fig:STO}(c), this film has a 4.6~nm interfacial region with reduced magnetization. The average interfacial magnetization is $47 \pm 5 \%$ and $46 \pm 5 \%$ of that of the bulk magnetization at 300~K and 120~K, respectively. 

Figure~\ref{Fig:STObuff} shows the results from the 20-nm SCRO film on a 50-nm SCNO buffer layer on STO. As expected, the frequency of the oscillations is higher in the XRR and PNR data due to a larger total thickness.  XRR data show a conspicuous beating effect from the contrast in the x-ray SLDs between SCRO and SCNO.  The nuclear scattering densities between SCRO and SCNO are very close, thus PNR data barely display any beating effect. Similar to the case of the SCRO on LSAT, the weak contrast in the nuclear scattering densities between SCRO and SCNO is advantageous for determining the interfacial magnetization structure. A layer with suppressed magnetization that is 3.6~nm thick is resolved at the interface between the SCRO and SCNO. The magnetization of this layer is $50\pm4$\% and $56\pm5$\% of the bulk magnetization at 300 K and 120 K, respectively.  Thus, with a SCNO buffer layer, the thickness of the suppressed magnetization layer is reduced, but the relative magnetization reduction is comparable to that of the SCRO film grown directly on STO.

Provided that the interfacial magnetization reduction is caused by anti-site disorders, the degree of the Cr/Re disorder can be estimated from the magnetization reduction. In the double exchange model for double perovskites ~\cite{vaitheeswaran2005pseudo, meetei2013theory, das2011origin}, the Cr 3$d$ and Re 5$d$ orbitals hybridize via the ligand oxygen orbitals, and there is an antiferromagnetic alignment between the rock-salt ordered Cr and Re spin moments. We assume that Cr carries 3~$\mu_B$ per ion and Re carries 2~$\mu_B$ per ion. Using a simple argument based on the ferrimagnetic model~\cite{serrate2007double},  each misplaced Cr ion reduces the saturation magnetization $M_{S}$ by 2 $\times$ 3 = 6 $\mu_B$, and each misplaced Re ion increases $M_{S}$ by $2 \times 2 = 4$~$\mu_B$. Thus, $m_{disordered}/m_{ordered} = 1-2x$, where $x$ is the percentage of the Cr/Re disorder.  From the relative average magnetization reduction at 300~K and 120~K, we found that the interfacial Cr/Re antisite disorder are comparable for the two films directly grown on LSAT and STO substrates, which are $22 \pm 5\%$ and $27 \pm 3\%$, respectively.  At the same time, the thicknesses of the magnetization suppression regions are also similar, although the two films are under different degrees of strain. Therefore, strain is unlikely the main cause for suppressed magnetization and the Cr/Re antisite disorder at interfaces. 

With a SCNO buffer layer, the thickness of the region with decreased magnetization reduces to 3.6~nm. Based on the aforementioned assumption, the average Cr/Re antisite disorder in this region is of $27 \pm 3\%$. The magnitude of the magnetization suppression in the interfacial regions is comparable to the films without a buffer layer. Interestingly, earlier STEM work found an interfacial disorder region of about 1-2~nm~\cite{lucy2013buffer}, which is less than the thickness with suppressed magnetization revealed by PNR.  The discrepancy may be related to the different lateral length scales of the two methods. PNR probes the average magnetization over the whole film plane with a lateral length scale of 10~mm,  while STEM probes a much smaller area, with a characteristic length of only 10~nm.  On the other hand, the two techniques probe two different physical quantities. Beside the antisite disorders, several other mechanisms, such as oxygen deficiency~\cite{hauser2013electronic, schumacher2013inducing} and inter-diffusion~\cite{kourkoutis2010microscopic},  can affect the magnetization of magnetic oxides. Particularly, it has been shown that the magnetic properties of SCRO films are very sensitive to the oxygen partial pressures during growth~\cite{hauser2013electronic}. Thus, it is of interest to determine whether the interfacial region and bulk part have a different level of oxygen deficiency, which can give rise to the difference between the thickness of the suppressed magnetization and the thickness of the antisite disorder region. Perhaps more interestingly, the discrepancy can also arise from interfacial electronic reconstruction. Interfacial electronic reconstruction can change the valence state and/or the orbital occupation of the transition metal ions,  thus affecting the magnetic properties at interfaces, which has been frequently observed in manganites~\cite{hoffmann2005suppressed, kavich2007nanoscale}.  Overall, current experiments focus on the effect of strain and buffer layer and are not able to pin down the leading sources, which is worth further investigation.    

In summary, we have studied the depth profiles of the magnetization structures of a series of 20-nm SCRO epitaxial films on LSAT and STO substrates, and on a 50-nm SCNO buffer layer on STO. These films shows magnetization suppression at interfaces with the substrate or the buffer, which is correlated with the B/B' antisite disorder. The two films deposited directly on LSAT and STO substrates show a 5-6~nm interfacial region with reduced magnetization, which corresponds to the Cr/Re antisite disorder of $22 \pm 5\%$ and $27 \pm 3 \%$ on LSAT and STO, respectively. For the SCRO film grown on a SCNO buffer layer on STO, the thickness of the magnetization suppression region is reduced to 3.6~nm with an average Cr/Re antisite disorder of  $27 \pm 3\%$.  However, the magnetization suppression region is thicker than the Cr/Re antisite disorder region at the interface between SCRO and SCNO,  which suggests that there exist other mechanisms for interfacial magnetization suppression.

\begin{acknowledgments}
We thank R. Goyette (ORNL) for assistance in PNR experiments and J. Pearson (ANL) for use of the SQUID magnetometer.  Work at Argonne National Laboratory (YL and StV) was supported by the U.S. Department of Energy (DOE), Office of Science, Basic Energy Sciences (BES), Materials Sciences and Engineering Division. Research at OSU was supported by the Center for Emergent Materials at the Ohio State University, a NSF Materials Research Science and Engineering Center (DMR-1420451). Research conducted at ORNL's Spallation Neutron Source was sponsored by the Scientific User Facilities Division, Office of Basic Energy Sciences, US Department of Energy.
\end{acknowledgments}

\section{APPENDIX}
\begin{figure}[h]
	\centering
		\includegraphics[width=0.448\textwidth]{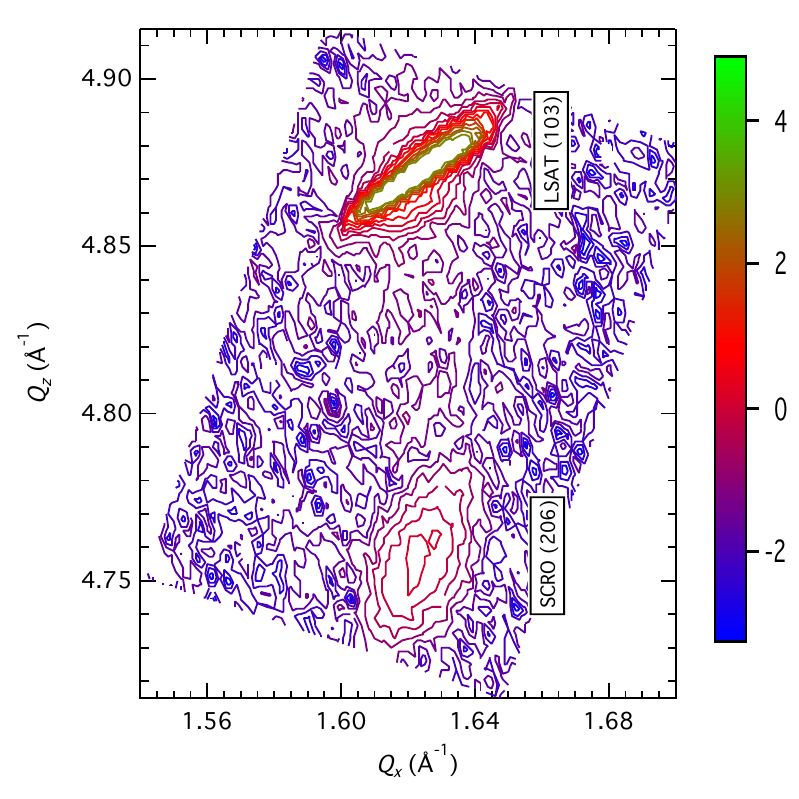}
	\caption{\label{Fig:RSM}(Color online) Reciprocal space map of the 20 nm thick SCRO film around the (103) LSAT substrate peak, which shows that the film is fully strained within experimental resolution.} 
\end{figure}

\bibliography{SCRO}	

\begin{thebibliography}{34}%
\makeatletter
\providecommand \@ifxundefined [1]{%
 \@ifx{#1\undefined}
}%
\providecommand \@ifnum [1]{%
 \ifnum #1\expandafter \@firstoftwo
 \else \expandafter \@secondoftwo
 \fi
}%
\providecommand \@ifx [1]{%
 \ifx #1\expandafter \@firstoftwo
 \else \expandafter \@secondoftwo
 \fi
}%
\providecommand \natexlab [1]{#1}%
\providecommand \enquote  [1]{``#1''}%
\providecommand \bibnamefont  [1]{#1}%
\providecommand \bibfnamefont [1]{#1}%
\providecommand \citenamefont [1]{#1}%
\providecommand \href@noop [0]{\@secondoftwo}%
\providecommand \href [0]{\begingroup \@sanitize@url \@href}%
\providecommand \@href[1]{\@@startlink{#1}\@@href}%
\providecommand \@@href[1]{\endgroup#1\@@endlink}%
\providecommand \@sanitize@url [0]{\catcode `\\12\catcode `\$12\catcode
  `\&12\catcode `\#12\catcode `\^12\catcode `\_12\catcode `\%12\relax}%
\providecommand \@@startlink[1]{}%
\providecommand \@@endlink[0]{}%
\providecommand \url  [0]{\begingroup\@sanitize@url \@url }%
\providecommand \@url [1]{\endgroup\@href {#1}{\urlprefix }}%
\providecommand \urlprefix  [0]{URL }%
\providecommand \Eprint [0]{\href }%
\providecommand \doibase [0]{http://dx.doi.org/}%
\providecommand \selectlanguage [0]{\@gobble}%
\providecommand \bibinfo  [0]{\@secondoftwo}%
\providecommand \bibfield  [0]{\@secondoftwo}%
\providecommand \translation [1]{[#1]}%
\providecommand \BibitemOpen [0]{}%
\providecommand \bibitemStop [0]{}%
\providecommand \bibitemNoStop [0]{.\EOS\space}%
\providecommand \EOS [0]{\spacefactor3000\relax}%
\providecommand \BibitemShut  [1]{\csname bibitem#1\endcsname}%
\let\auto@bib@innerbib\@empty
\bibitem [{\citenamefont {Serrate}\ \emph {et~al.}(2007)\citenamefont
  {Serrate}, \citenamefont {De~Teresa},\ and\ \citenamefont
  {Ibarra}}]{serrate2007double}%
  \BibitemOpen
  \bibfield  {author} {\bibinfo {author} {\bibfnamefont {D.}~\bibnamefont
  {Serrate}}, \bibinfo {author} {\bibfnamefont {J.}~\bibnamefont {De~Teresa}},
  \ and\ \bibinfo {author} {\bibfnamefont {M.}~\bibnamefont {Ibarra}},\
  }\href@noop {} {\bibfield  {journal} {\bibinfo  {journal} {Journal of
  Physics: Condensed Matter}\ }\textbf {\bibinfo {volume} {19}},\ \bibinfo
  {pages} {023201} (\bibinfo {year} {2007})}\BibitemShut {NoStop}%
\bibitem [{\citenamefont {Kato}\ \emph {et~al.}(2002)\citenamefont {Kato},
  \citenamefont {Okuda}, \citenamefont {Okimoto}, \citenamefont {Tomioka},
  \citenamefont {Takenoya}, \citenamefont {Ohkubo}, \citenamefont {Kawasaki},\
  and\ \citenamefont {Tokura}}]{kato2002metallic}%
  \BibitemOpen
  \bibfield  {author} {\bibinfo {author} {\bibfnamefont {H.}~\bibnamefont
  {Kato}}, \bibinfo {author} {\bibfnamefont {T.}~\bibnamefont {Okuda}},
  \bibinfo {author} {\bibfnamefont {Y.}~\bibnamefont {Okimoto}}, \bibinfo
  {author} {\bibfnamefont {Y.}~\bibnamefont {Tomioka}}, \bibinfo {author}
  {\bibfnamefont {Y.}~\bibnamefont {Takenoya}}, \bibinfo {author}
  {\bibfnamefont {A.}~\bibnamefont {Ohkubo}}, \bibinfo {author} {\bibfnamefont
  {M.}~\bibnamefont {Kawasaki}}, \ and\ \bibinfo {author} {\bibfnamefont
  {Y.}~\bibnamefont {Tokura}},\ }\href@noop {} {\bibfield  {journal} {\bibinfo
  {journal} {Applied physics letters}\ }\textbf {\bibinfo {volume} {81}},\
  \bibinfo {pages} {328} (\bibinfo {year} {2002})}\BibitemShut {NoStop}%
\bibitem [{\citenamefont {De~Teresa}\ \emph {et~al.}(2005)\citenamefont
  {De~Teresa}, \citenamefont {Serrate}, \citenamefont {Ritter}, \citenamefont
  {Blasco}, \citenamefont {Ibarra}, \citenamefont {Morellon},\ and\
  \citenamefont {Tokarz}}]{de2005investigation}%
  \BibitemOpen
  \bibfield  {author} {\bibinfo {author} {\bibfnamefont {J.}~\bibnamefont
  {De~Teresa}}, \bibinfo {author} {\bibfnamefont {D.}~\bibnamefont {Serrate}},
  \bibinfo {author} {\bibfnamefont {C.}~\bibnamefont {Ritter}}, \bibinfo
  {author} {\bibfnamefont {J.}~\bibnamefont {Blasco}}, \bibinfo {author}
  {\bibfnamefont {M.}~\bibnamefont {Ibarra}}, \bibinfo {author} {\bibfnamefont
  {L.}~\bibnamefont {Morellon}}, \ and\ \bibinfo {author} {\bibfnamefont
  {W.}~\bibnamefont {Tokarz}},\ }\href@noop {} {\bibfield  {journal} {\bibinfo
  {journal} {Physical Review B}\ }\textbf {\bibinfo {volume} {71}},\ \bibinfo
  {pages} {092408} (\bibinfo {year} {2005})}\BibitemShut {NoStop}%
\bibitem [{\citenamefont {Hauser}\ \emph {et~al.}(2012)\citenamefont {Hauser},
  \citenamefont {Soliz}, \citenamefont {Dixit}, \citenamefont {Williams},
  \citenamefont {Susner}, \citenamefont {Peters}, \citenamefont {Mier},
  \citenamefont {Gustafson}, \citenamefont {Sumption}, \citenamefont {Fraser}
  \emph {et~al.}}]{hauser2012fully}%
  \BibitemOpen
  \bibfield  {author} {\bibinfo {author} {\bibfnamefont {A.}~\bibnamefont
  {Hauser}}, \bibinfo {author} {\bibfnamefont {J.}~\bibnamefont {Soliz}},
  \bibinfo {author} {\bibfnamefont {M.}~\bibnamefont {Dixit}}, \bibinfo
  {author} {\bibfnamefont {R.}~\bibnamefont {Williams}}, \bibinfo {author}
  {\bibfnamefont {M.}~\bibnamefont {Susner}}, \bibinfo {author} {\bibfnamefont
  {B.}~\bibnamefont {Peters}}, \bibinfo {author} {\bibfnamefont
  {L.}~\bibnamefont {Mier}}, \bibinfo {author} {\bibfnamefont {T.}~\bibnamefont
  {Gustafson}}, \bibinfo {author} {\bibfnamefont {M.}~\bibnamefont {Sumption}},
  \bibinfo {author} {\bibfnamefont {H.}~\bibnamefont {Fraser}},  \emph
  {et~al.},\ }\href@noop {} {\bibfield  {journal} {\bibinfo  {journal}
  {Physical Review B}\ }\textbf {\bibinfo {volume} {85}},\ \bibinfo {pages}
  {161201} (\bibinfo {year} {2012})}\BibitemShut {NoStop}%
\bibitem [{\citenamefont {Jeon}\ \emph {et~al.}(2014)\citenamefont {Jeon},
  \citenamefont {Min}, \citenamefont {Spiesser}, \citenamefont {Saito},
  \citenamefont {Shin}, \citenamefont {Yuasa},\ and\ \citenamefont
  {Jansen}}]{jeon2014voltage}%
  \BibitemOpen
  \bibfield  {author} {\bibinfo {author} {\bibfnamefont {K.-R.}\ \bibnamefont
  {Jeon}}, \bibinfo {author} {\bibfnamefont {B.-C.}\ \bibnamefont {Min}},
  \bibinfo {author} {\bibfnamefont {A.}~\bibnamefont {Spiesser}}, \bibinfo
  {author} {\bibfnamefont {H.}~\bibnamefont {Saito}}, \bibinfo {author}
  {\bibfnamefont {S.-C.}\ \bibnamefont {Shin}}, \bibinfo {author}
  {\bibfnamefont {S.}~\bibnamefont {Yuasa}}, \ and\ \bibinfo {author}
  {\bibfnamefont {R.}~\bibnamefont {Jansen}},\ }\href@noop {} {\bibfield
  {journal} {\bibinfo  {journal} {Nature materials}\ }\textbf {\bibinfo
  {volume} {13}},\ \bibinfo {pages} {360} (\bibinfo {year} {2014})}\BibitemShut
  {NoStop}%
\bibitem [{\citenamefont {Van't~Erve}\ \emph {et~al.}(2012)\citenamefont
  {Van't~Erve}, \citenamefont {Friedman}, \citenamefont {Cobas}, \citenamefont
  {Li}, \citenamefont {Robinson},\ and\ \citenamefont {Jonker}}]{van2012low}%
  \BibitemOpen
  \bibfield  {author} {\bibinfo {author} {\bibfnamefont {O.}~\bibnamefont
  {Van't~Erve}}, \bibinfo {author} {\bibfnamefont {A.}~\bibnamefont
  {Friedman}}, \bibinfo {author} {\bibfnamefont {E.}~\bibnamefont {Cobas}},
  \bibinfo {author} {\bibfnamefont {C.}~\bibnamefont {Li}}, \bibinfo {author}
  {\bibfnamefont {J.}~\bibnamefont {Robinson}}, \ and\ \bibinfo {author}
  {\bibfnamefont {B.}~\bibnamefont {Jonker}},\ }\href@noop {} {\bibfield
  {journal} {\bibinfo  {journal} {Nature nanotechnology}\ }\textbf {\bibinfo
  {volume} {7}},\ \bibinfo {pages} {737} (\bibinfo {year} {2012})}\BibitemShut
  {NoStop}%
\bibitem [{\citenamefont {Raman}\ \emph {et~al.}(2011)\citenamefont {Raman},
  \citenamefont {Chang},\ and\ \citenamefont {Moodera}}]{raman2011new}%
  \BibitemOpen
  \bibfield  {author} {\bibinfo {author} {\bibfnamefont {K.}~\bibnamefont
  {Raman}}, \bibinfo {author} {\bibfnamefont {J.}~\bibnamefont {Chang}}, \ and\
  \bibinfo {author} {\bibfnamefont {J.}~\bibnamefont {Moodera}},\ }\href@noop
  {} {\bibfield  {journal} {\bibinfo  {journal} {Organic Electronics}\ }\textbf
  {\bibinfo {volume} {12}},\ \bibinfo {pages} {1275} (\bibinfo {year}
  {2011})}\BibitemShut {NoStop}%
\bibitem [{\citenamefont {Barraud}\ \emph {et~al.}(2010)\citenamefont
  {Barraud}, \citenamefont {Seneor}, \citenamefont {Mattana}, \citenamefont
  {Fusil}, \citenamefont {Bouzehouane}, \citenamefont {Deranlot}, \citenamefont
  {Graziosi}, \citenamefont {Hueso}, \citenamefont {Bergenti}, \citenamefont
  {Dediu} \emph {et~al.}}]{barraud2010unravelling}%
  \BibitemOpen
  \bibfield  {author} {\bibinfo {author} {\bibfnamefont {C.}~\bibnamefont
  {Barraud}}, \bibinfo {author} {\bibfnamefont {P.}~\bibnamefont {Seneor}},
  \bibinfo {author} {\bibfnamefont {R.}~\bibnamefont {Mattana}}, \bibinfo
  {author} {\bibfnamefont {S.}~\bibnamefont {Fusil}}, \bibinfo {author}
  {\bibfnamefont {K.}~\bibnamefont {Bouzehouane}}, \bibinfo {author}
  {\bibfnamefont {C.}~\bibnamefont {Deranlot}}, \bibinfo {author}
  {\bibfnamefont {P.}~\bibnamefont {Graziosi}}, \bibinfo {author}
  {\bibfnamefont {L.}~\bibnamefont {Hueso}}, \bibinfo {author} {\bibfnamefont
  {I.}~\bibnamefont {Bergenti}}, \bibinfo {author} {\bibfnamefont
  {V.}~\bibnamefont {Dediu}},  \emph {et~al.},\ }\href@noop {} {\bibfield
  {journal} {\bibinfo  {journal} {Nature Physics}\ }\textbf {\bibinfo {volume}
  {6}},\ \bibinfo {pages} {615} (\bibinfo {year} {2010})}\BibitemShut {NoStop}%
\bibitem [{\citenamefont {Datta}\ and\ \citenamefont
  {Das}(1990)}]{datta1990electronic}%
  \BibitemOpen
  \bibfield  {author} {\bibinfo {author} {\bibfnamefont {S.}~\bibnamefont
  {Datta}}\ and\ \bibinfo {author} {\bibfnamefont {B.}~\bibnamefont {Das}},\
  }\href@noop {} {\bibfield  {journal} {\bibinfo  {journal} {Applied Physics
  Letters}\ }\textbf {\bibinfo {volume} {56}},\ \bibinfo {pages} {665}
  (\bibinfo {year} {1990})}\BibitemShut {NoStop}%
\bibitem [{\citenamefont {Rashba}(1960)}]{rashba1960properties}%
  \BibitemOpen
  \bibfield  {author} {\bibinfo {author} {\bibfnamefont {E.}~\bibnamefont
  {Rashba}},\ }\href@noop {} {\bibfield  {journal} {\bibinfo  {journal} {Soviet
  Physics-Solid State}\ }\textbf {\bibinfo {volume} {2}},\ \bibinfo {pages}
  {1109} (\bibinfo {year} {1960})}\BibitemShut {NoStop}%
\bibitem [{\citenamefont {Schmidt}\ \emph {et~al.}(2000)\citenamefont
  {Schmidt}, \citenamefont {Ferrand}, \citenamefont {Molenkamp}, \citenamefont
  {Filip},\ and\ \citenamefont {Van~Wees}}]{schmidt2000fundamental}%
  \BibitemOpen
  \bibfield  {author} {\bibinfo {author} {\bibfnamefont {G.}~\bibnamefont
  {Schmidt}}, \bibinfo {author} {\bibfnamefont {D.}~\bibnamefont {Ferrand}},
  \bibinfo {author} {\bibfnamefont {L.}~\bibnamefont {Molenkamp}}, \bibinfo
  {author} {\bibfnamefont {A.}~\bibnamefont {Filip}}, \ and\ \bibinfo {author}
  {\bibfnamefont {B.}~\bibnamefont {Van~Wees}},\ }\href@noop {} {\bibfield
  {journal} {\bibinfo  {journal} {Physical Review B}\ }\textbf {\bibinfo
  {volume} {62}},\ \bibinfo {pages} {R4790} (\bibinfo {year}
  {2000})}\BibitemShut {NoStop}%
\bibitem [{\citenamefont {Smith}\ and\ \citenamefont
  {Silver}(2001)}]{smith2001electrical}%
  \BibitemOpen
  \bibfield  {author} {\bibinfo {author} {\bibfnamefont {D.}~\bibnamefont
  {Smith}}\ and\ \bibinfo {author} {\bibfnamefont {R.}~\bibnamefont {Silver}},\
  }\href@noop {} {\bibfield  {journal} {\bibinfo  {journal} {Physical review
  B}\ }\textbf {\bibinfo {volume} {64}},\ \bibinfo {pages} {045323} (\bibinfo
  {year} {2001})}\BibitemShut {NoStop}%
\bibitem [{\citenamefont {Komelj}(2010)}]{komelj2010magnetoelasticity}%
  \BibitemOpen
  \bibfield  {author} {\bibinfo {author} {\bibfnamefont {M.}~\bibnamefont
  {Komelj}},\ }\href@noop {} {\bibfield  {journal} {\bibinfo  {journal}
  {Physical Review B}\ }\textbf {\bibinfo {volume} {82}},\ \bibinfo {pages}
  {012410} (\bibinfo {year} {2010})}\BibitemShut {NoStop}%
\bibitem [{\citenamefont {Czeschka}\ \emph {et~al.}(2009)\citenamefont
  {Czeschka}, \citenamefont {Gepraegs}, \citenamefont {Opel}, \citenamefont
  {Goennenwein},\ and\ \citenamefont {Gross}}]{czeschka2009giant}%
  \BibitemOpen
  \bibfield  {author} {\bibinfo {author} {\bibfnamefont {F.~D.}\ \bibnamefont
  {Czeschka}}, \bibinfo {author} {\bibfnamefont {S.}~\bibnamefont {Gepraegs}},
  \bibinfo {author} {\bibfnamefont {M.}~\bibnamefont {Opel}}, \bibinfo {author}
  {\bibfnamefont {S.~T.}\ \bibnamefont {Goennenwein}}, \ and\ \bibinfo {author}
  {\bibfnamefont {R.}~\bibnamefont {Gross}},\ }\href@noop {} {\bibfield
  {journal} {\bibinfo  {journal} {Applied Physics Letters}\ }\textbf {\bibinfo
  {volume} {95}},\ \bibinfo {pages} {062508} (\bibinfo {year}
  {2009})}\BibitemShut {NoStop}%
\bibitem [{\citenamefont {Hauser}\ \emph {et~al.}(2013)\citenamefont {Hauser},
  \citenamefont {Lucy}, \citenamefont {Wang}, \citenamefont {Soliz},
  \citenamefont {Holcomb}, \citenamefont {Morris}, \citenamefont {Woodward},\
  and\ \citenamefont {Yang}}]{hauser2013electronic}%
  \BibitemOpen
  \bibfield  {author} {\bibinfo {author} {\bibfnamefont {A.}~\bibnamefont
  {Hauser}}, \bibinfo {author} {\bibfnamefont {J.}~\bibnamefont {Lucy}},
  \bibinfo {author} {\bibfnamefont {H.}~\bibnamefont {Wang}}, \bibinfo {author}
  {\bibfnamefont {J.}~\bibnamefont {Soliz}}, \bibinfo {author} {\bibfnamefont
  {A.}~\bibnamefont {Holcomb}}, \bibinfo {author} {\bibfnamefont
  {P.}~\bibnamefont {Morris}}, \bibinfo {author} {\bibfnamefont
  {P.}~\bibnamefont {Woodward}}, \ and\ \bibinfo {author} {\bibfnamefont
  {F.}~\bibnamefont {Yang}},\ }\href@noop {} {\bibfield  {journal} {\bibinfo
  {journal} {Applied Physics Letters}\ }\textbf {\bibinfo {volume} {102}},\
  \bibinfo {pages} {032403} (\bibinfo {year} {2013})}\BibitemShut {NoStop}%
\bibitem [{\citenamefont {Lucy}\ \emph {et~al.}(2013)\citenamefont {Lucy},
  \citenamefont {Hauser}, \citenamefont {Wang}, \citenamefont {Soliz},
  \citenamefont {Dixit}, \citenamefont {Williams}, \citenamefont {Holcombe},
  \citenamefont {Morris}, \citenamefont {Fraser}, \citenamefont {McComb} \emph
  {et~al.}}]{lucy2013buffer}%
  \BibitemOpen
  \bibfield  {author} {\bibinfo {author} {\bibfnamefont {J.}~\bibnamefont
  {Lucy}}, \bibinfo {author} {\bibfnamefont {A.}~\bibnamefont {Hauser}},
  \bibinfo {author} {\bibfnamefont {H.}~\bibnamefont {Wang}}, \bibinfo {author}
  {\bibfnamefont {J.}~\bibnamefont {Soliz}}, \bibinfo {author} {\bibfnamefont
  {M.}~\bibnamefont {Dixit}}, \bibinfo {author} {\bibfnamefont
  {R.}~\bibnamefont {Williams}}, \bibinfo {author} {\bibfnamefont
  {A.}~\bibnamefont {Holcombe}}, \bibinfo {author} {\bibfnamefont
  {P.}~\bibnamefont {Morris}}, \bibinfo {author} {\bibfnamefont
  {H.}~\bibnamefont {Fraser}}, \bibinfo {author} {\bibfnamefont
  {D.}~\bibnamefont {McComb}},  \emph {et~al.},\ }\href@noop {} {\bibfield
  {journal} {\bibinfo  {journal} {Applied Physics Letters}\ }\textbf {\bibinfo
  {volume} {103}},\ \bibinfo {pages} {042414} (\bibinfo {year}
  {2013})}\BibitemShut {NoStop}%
\bibitem [{\citenamefont {Tsymbal}\ \emph {et~al.}(2007)\citenamefont
  {Tsymbal}, \citenamefont {Belashchenko}, \citenamefont {Velev}, \citenamefont
  {Jaswal}, \citenamefont {Van~Schilfgaarde}, \citenamefont {Oleynik},\ and\
  \citenamefont {Stewart}}]{tsymbal2007interface}%
  \BibitemOpen
  \bibfield  {author} {\bibinfo {author} {\bibfnamefont {E.~Y.}\ \bibnamefont
  {Tsymbal}}, \bibinfo {author} {\bibfnamefont {K.~D.}\ \bibnamefont
  {Belashchenko}}, \bibinfo {author} {\bibfnamefont {J.~P.}\ \bibnamefont
  {Velev}}, \bibinfo {author} {\bibfnamefont {S.}~\bibnamefont {Jaswal}},
  \bibinfo {author} {\bibfnamefont {M.}~\bibnamefont {Van~Schilfgaarde}},
  \bibinfo {author} {\bibfnamefont {I.}~\bibnamefont {Oleynik}}, \ and\
  \bibinfo {author} {\bibfnamefont {D.}~\bibnamefont {Stewart}},\ }\href@noop
  {} {\bibfield  {journal} {\bibinfo  {journal} {Progress in materials
  science}\ }\textbf {\bibinfo {volume} {52}},\ \bibinfo {pages} {401}
  (\bibinfo {year} {2007})}\BibitemShut {NoStop}%
\bibitem [{\citenamefont {Liu}\ \emph {et~al.}(2009)\citenamefont {Liu},
  \citenamefont {Watson}, \citenamefont {Lee}, \citenamefont {Gorham},
  \citenamefont {Katz}, \citenamefont {Borchers}, \citenamefont {Fairbrother},\
  and\ \citenamefont {Reich}}]{liu2009correlation}%
  \BibitemOpen
  \bibfield  {author} {\bibinfo {author} {\bibfnamefont {Y.~H.}\ \bibnamefont
  {Liu}}, \bibinfo {author} {\bibfnamefont {S.~M.}\ \bibnamefont {Watson}},
  \bibinfo {author} {\bibfnamefont {T.}~\bibnamefont {Lee}}, \bibinfo {author}
  {\bibfnamefont {J.~M.}\ \bibnamefont {Gorham}}, \bibinfo {author}
  {\bibfnamefont {H.~E.}\ \bibnamefont {Katz}}, \bibinfo {author}
  {\bibfnamefont {J.~A.}\ \bibnamefont {Borchers}}, \bibinfo {author}
  {\bibfnamefont {H.~D.}\ \bibnamefont {Fairbrother}}, \ and\ \bibinfo {author}
  {\bibfnamefont {D.~H.}\ \bibnamefont {Reich}},\ }\href@noop {} {\bibfield
  {journal} {\bibinfo  {journal} {Physical Review B}\ }\textbf {\bibinfo
  {volume} {79}},\ \bibinfo {pages} {075312} (\bibinfo {year}
  {2009})}\BibitemShut {NoStop}%
\bibitem [{\citenamefont {Liu}\ \emph {et~al.}(2013)\citenamefont {Liu},
  \citenamefont {Cuellar}, \citenamefont {Sefrioui}, \citenamefont {Freeland},
  \citenamefont {Fitzsimmons}, \citenamefont {Leon}, \citenamefont
  {Santamaria},\ and\ \citenamefont {te~Velthuis}}]{liu2013emergent}%
  \BibitemOpen
  \bibfield  {author} {\bibinfo {author} {\bibfnamefont {Y.~H.}\ \bibnamefont
  {Liu}}, \bibinfo {author} {\bibfnamefont {F.}~\bibnamefont {Cuellar}},
  \bibinfo {author} {\bibfnamefont {Z.}~\bibnamefont {Sefrioui}}, \bibinfo
  {author} {\bibfnamefont {J.}~\bibnamefont {Freeland}}, \bibinfo {author}
  {\bibfnamefont {M.}~\bibnamefont {Fitzsimmons}}, \bibinfo {author}
  {\bibfnamefont {C.}~\bibnamefont {Leon}}, \bibinfo {author} {\bibfnamefont
  {J.}~\bibnamefont {Santamaria}}, \ and\ \bibinfo {author} {\bibfnamefont
  {S.~G.~E.}\ \bibnamefont {te~Velthuis}},\ }\href@noop {} {\bibfield
  {journal} {\bibinfo  {journal} {Physical Review Letters}\ }\textbf {\bibinfo
  {volume} {111}},\ \bibinfo {pages} {247203} (\bibinfo {year}
  {2013})}\BibitemShut {NoStop}%
\bibitem [{\citenamefont {Felcher}\ \emph {et~al.}(1987)\citenamefont
  {Felcher}, \citenamefont {Hilleke}, \citenamefont {Crawford}, \citenamefont
  {Haumann}, \citenamefont {Kleb},\ and\ \citenamefont
  {Ostrowski}}]{FelcherRSI1987}%
  \BibitemOpen
  \bibfield  {author} {\bibinfo {author} {\bibfnamefont {G.~P.}\ \bibnamefont
  {Felcher}}, \bibinfo {author} {\bibfnamefont {R.~O.}\ \bibnamefont
  {Hilleke}}, \bibinfo {author} {\bibfnamefont {R.~K.}\ \bibnamefont
  {Crawford}}, \bibinfo {author} {\bibfnamefont {J.}~\bibnamefont {Haumann}},
  \bibinfo {author} {\bibfnamefont {R.}~\bibnamefont {Kleb}}, \ and\ \bibinfo
  {author} {\bibfnamefont {G.}~\bibnamefont {Ostrowski}},\ }\href {\doibase
  10.1063/1.1139225} {\bibfield  {journal} {\bibinfo  {journal} {Rev. Sci.
  Instrum.}\ }\textbf {\bibinfo {volume} {58}},\ \bibinfo {pages} {609}
  (\bibinfo {year} {1987})}\BibitemShut {NoStop}%
\bibitem [{\citenamefont {Majkrzak}(1996)}]{MajkrzakPhysB1996}%
  \BibitemOpen
  \bibfield  {author} {\bibinfo {author} {\bibfnamefont {C.}~\bibnamefont
  {Majkrzak}},\ }\href@noop {} {\bibfield  {journal} {\bibinfo  {journal}
  {Physica B: Condensed Matter}\ }\textbf {\bibinfo {volume} {221}},\ \bibinfo
  {pages} {342} (\bibinfo {year} {1996})}\BibitemShut {NoStop}%
\bibitem [{\citenamefont {Cowley}\ \emph {et~al.}(1969)\citenamefont {Cowley},
  \citenamefont {Buyers},\ and\ \citenamefont
  {Dolling}}]{cowley1969relationship}%
  \BibitemOpen
  \bibfield  {author} {\bibinfo {author} {\bibfnamefont {R.}~\bibnamefont
  {Cowley}}, \bibinfo {author} {\bibfnamefont {W.}~\bibnamefont {Buyers}}, \
  and\ \bibinfo {author} {\bibfnamefont {G.}~\bibnamefont {Dolling}},\
  }\href@noop {} {\bibfield  {journal} {\bibinfo  {journal} {Solid State
  Communications}\ }\textbf {\bibinfo {volume} {7}},\ \bibinfo {pages} {181}
  (\bibinfo {year} {1969})}\BibitemShut {NoStop}%
\bibitem [{\citenamefont {Parratt}(1954)}]{ParrattPR1954}%
  \BibitemOpen
  \bibfield  {author} {\bibinfo {author} {\bibfnamefont {L.~G.}\ \bibnamefont
  {Parratt}},\ }\href {\doibase 10.1103/PhysRev.95.359} {\bibfield  {journal}
  {\bibinfo  {journal} {Phys. Rev.}\ }\textbf {\bibinfo {volume} {95}},\
  \bibinfo {pages} {359} (\bibinfo {year} {1954})}\BibitemShut {NoStop}%
\bibitem [{\citenamefont {Liu}\ \emph {et~al.}(2011)\citenamefont {Liu},
  \citenamefont {te~Velthuis}, \citenamefont {Jiang}, \citenamefont {Choi},
  \citenamefont {Bader}, \citenamefont {Parizzi}, \citenamefont {Ambaye},\ and\
  \citenamefont {Lauter}}]{liu2011magnetic}%
  \BibitemOpen
  \bibfield  {author} {\bibinfo {author} {\bibfnamefont {Y.~H.}\ \bibnamefont
  {Liu}}, \bibinfo {author} {\bibfnamefont {S.~G.~E.}\ \bibnamefont
  {te~Velthuis}}, \bibinfo {author} {\bibfnamefont {J.}~\bibnamefont {Jiang}},
  \bibinfo {author} {\bibfnamefont {Y.}~\bibnamefont {Choi}}, \bibinfo {author}
  {\bibfnamefont {S.}~\bibnamefont {Bader}}, \bibinfo {author} {\bibfnamefont
  {A.}~\bibnamefont {Parizzi}}, \bibinfo {author} {\bibfnamefont
  {H.}~\bibnamefont {Ambaye}}, \ and\ \bibinfo {author} {\bibfnamefont
  {V.}~\bibnamefont {Lauter}},\ }\href@noop {} {\bibfield  {journal} {\bibinfo
  {journal} {Physical Review B}\ }\textbf {\bibinfo {volume} {83}},\ \bibinfo
  {pages} {174418} (\bibinfo {year} {2011})}\BibitemShut {NoStop}%
\bibitem [{IGO()}]{IGORPro}%
  \BibitemOpen
  \href@noop {} {}\bibinfo {note}
  {Http://www.wavemetrics.com/products/igorpro/\-dataanalysis/curvefitting.htm.}\BibitemShut
  {Stop}%
\bibitem [{\citenamefont {Rutkowski}\ \emph {et~al.}(2010)\citenamefont
  {Rutkowski}, \citenamefont {Hauser}, \citenamefont {Yang}, \citenamefont
  {Ricciardo}, \citenamefont {Meyer}, \citenamefont {Woodward}, \citenamefont
  {Holcombe}, \citenamefont {Morris},\ and\ \citenamefont
  {Brillson}}]{rutkowski2010x}%
  \BibitemOpen
  \bibfield  {author} {\bibinfo {author} {\bibfnamefont {M.}~\bibnamefont
  {Rutkowski}}, \bibinfo {author} {\bibfnamefont {A.}~\bibnamefont {Hauser}},
  \bibinfo {author} {\bibfnamefont {F.}~\bibnamefont {Yang}}, \bibinfo {author}
  {\bibfnamefont {R.}~\bibnamefont {Ricciardo}}, \bibinfo {author}
  {\bibfnamefont {T.}~\bibnamefont {Meyer}}, \bibinfo {author} {\bibfnamefont
  {P.}~\bibnamefont {Woodward}}, \bibinfo {author} {\bibfnamefont
  {A.}~\bibnamefont {Holcombe}}, \bibinfo {author} {\bibfnamefont
  {P.}~\bibnamefont {Morris}}, \ and\ \bibinfo {author} {\bibfnamefont
  {L.}~\bibnamefont {Brillson}},\ }\href@noop {} {\bibfield  {journal}
  {\bibinfo  {journal} {Journal of Vacuum Science \& Technology A}\ }\textbf
  {\bibinfo {volume} {28}},\ \bibinfo {pages} {1240} (\bibinfo {year}
  {2010})}\BibitemShut {NoStop}%
\bibitem [{\citenamefont {Jalili}\ \emph {et~al.}(2009)\citenamefont {Jalili},
  \citenamefont {Heinig},\ and\ \citenamefont {Leung}}]{jalili2009x}%
  \BibitemOpen
  \bibfield  {author} {\bibinfo {author} {\bibfnamefont {H.}~\bibnamefont
  {Jalili}}, \bibinfo {author} {\bibfnamefont {N.}~\bibnamefont {Heinig}}, \
  and\ \bibinfo {author} {\bibfnamefont {K.}~\bibnamefont {Leung}},\
  }\href@noop {} {\bibfield  {journal} {\bibinfo  {journal} {Physical Review
  B}\ }\textbf {\bibinfo {volume} {79}},\ \bibinfo {pages} {174427} (\bibinfo
  {year} {2009})}\BibitemShut {NoStop}%
\bibitem [{\citenamefont {Vaitheeswaran}\ \emph {et~al.}(2005)\citenamefont
  {Vaitheeswaran}, \citenamefont {Kanchana},\ and\ \citenamefont
  {Delin}}]{vaitheeswaran2005pseudo}%
  \BibitemOpen
  \bibfield  {author} {\bibinfo {author} {\bibfnamefont {G.}~\bibnamefont
  {Vaitheeswaran}}, \bibinfo {author} {\bibfnamefont {V.}~\bibnamefont
  {Kanchana}}, \ and\ \bibinfo {author} {\bibfnamefont {A.}~\bibnamefont
  {Delin}},\ }\href@noop {} {\bibfield  {journal} {\bibinfo  {journal} {Applied
  Physics Letters}\ }\textbf {\bibinfo {volume} {86}},\ \bibinfo {pages}
  {032513} (\bibinfo {year} {2005})}\BibitemShut {NoStop}%
\bibitem [{\citenamefont {Meetei}\ \emph {et~al.}(2013)\citenamefont {Meetei},
  \citenamefont {Erten}, \citenamefont {Mukherjee}, \citenamefont {Randeria},
  \citenamefont {Trivedi},\ and\ \citenamefont {Woodward}}]{meetei2013theory}%
  \BibitemOpen
  \bibfield  {author} {\bibinfo {author} {\bibfnamefont {O.~N.}\ \bibnamefont
  {Meetei}}, \bibinfo {author} {\bibfnamefont {O.}~\bibnamefont {Erten}},
  \bibinfo {author} {\bibfnamefont {A.}~\bibnamefont {Mukherjee}}, \bibinfo
  {author} {\bibfnamefont {M.}~\bibnamefont {Randeria}}, \bibinfo {author}
  {\bibfnamefont {N.}~\bibnamefont {Trivedi}}, \ and\ \bibinfo {author}
  {\bibfnamefont {P.}~\bibnamefont {Woodward}},\ }\href@noop {} {\bibfield
  {journal} {\bibinfo  {journal} {Physical Review B}\ }\textbf {\bibinfo
  {volume} {87}},\ \bibinfo {pages} {165104} (\bibinfo {year}
  {2013})}\BibitemShut {NoStop}%
\bibitem [{\citenamefont {Das}\ \emph {et~al.}(2011)\citenamefont {Das},
  \citenamefont {Sanyal}, \citenamefont {Saha-Dasgupta},\ and\ \citenamefont
  {Sarma}}]{das2011origin}%
  \BibitemOpen
  \bibfield  {author} {\bibinfo {author} {\bibfnamefont {H.}~\bibnamefont
  {Das}}, \bibinfo {author} {\bibfnamefont {P.}~\bibnamefont {Sanyal}},
  \bibinfo {author} {\bibfnamefont {T.}~\bibnamefont {Saha-Dasgupta}}, \ and\
  \bibinfo {author} {\bibfnamefont {D.}~\bibnamefont {Sarma}},\ }\href@noop {}
  {\bibfield  {journal} {\bibinfo  {journal} {Physical Review B}\ }\textbf
  {\bibinfo {volume} {83}},\ \bibinfo {pages} {104418} (\bibinfo {year}
  {2011})}\BibitemShut {NoStop}%
\bibitem [{\citenamefont {Schumacher}\ \emph {et~al.}(2013)\citenamefont
  {Schumacher}, \citenamefont {Steffen}, \citenamefont {Voigt}, \citenamefont
  {Schubert}, \citenamefont {Br{\"u}ckel}, \citenamefont {Ambaye},\ and\
  \citenamefont {Lauter}}]{schumacher2013inducing}%
  \BibitemOpen
  \bibfield  {author} {\bibinfo {author} {\bibfnamefont {D.}~\bibnamefont
  {Schumacher}}, \bibinfo {author} {\bibfnamefont {A.}~\bibnamefont {Steffen}},
  \bibinfo {author} {\bibfnamefont {J.}~\bibnamefont {Voigt}}, \bibinfo
  {author} {\bibfnamefont {J.}~\bibnamefont {Schubert}}, \bibinfo {author}
  {\bibfnamefont {T.}~\bibnamefont {Br{\"u}ckel}}, \bibinfo {author}
  {\bibfnamefont {H.}~\bibnamefont {Ambaye}}, \ and\ \bibinfo {author}
  {\bibfnamefont {V.}~\bibnamefont {Lauter}},\ }\href@noop {} {\bibfield
  {journal} {\bibinfo  {journal} {Physical Review B}\ }\textbf {\bibinfo
  {volume} {88}},\ \bibinfo {pages} {144427} (\bibinfo {year}
  {2013})}\BibitemShut {NoStop}%
\bibitem [{\citenamefont {Kourkoutis}\ \emph {et~al.}(2010)\citenamefont
  {Kourkoutis}, \citenamefont {Song}, \citenamefont {Hwang},\ and\
  \citenamefont {Muller}}]{kourkoutis2010microscopic}%
  \BibitemOpen
  \bibfield  {author} {\bibinfo {author} {\bibfnamefont {L.~F.}\ \bibnamefont
  {Kourkoutis}}, \bibinfo {author} {\bibfnamefont {J.}~\bibnamefont {Song}},
  \bibinfo {author} {\bibfnamefont {H.}~\bibnamefont {Hwang}}, \ and\ \bibinfo
  {author} {\bibfnamefont {D.}~\bibnamefont {Muller}},\ }\href@noop {}
  {\bibfield  {journal} {\bibinfo  {journal} {Proceedings of the National
  Academy of Sciences}\ }\textbf {\bibinfo {volume} {107}},\ \bibinfo {pages}
  {11682} (\bibinfo {year} {2010})}\BibitemShut {NoStop}%
\bibitem [{\citenamefont {Hoffmann}\ \emph {et~al.}(2005)\citenamefont
  {Hoffmann}, \citenamefont {te~Velthuis}, \citenamefont {Sefrioui},
  \citenamefont {Santamar{\'\i}a}, \citenamefont {Fitzsimmons}, \citenamefont
  {Park},\ and\ \citenamefont {Varela}}]{hoffmann2005suppressed}%
  \BibitemOpen
  \bibfield  {author} {\bibinfo {author} {\bibfnamefont {A.}~\bibnamefont
  {Hoffmann}}, \bibinfo {author} {\bibfnamefont {S.~G.~E.}\ \bibnamefont
  {te~Velthuis}}, \bibinfo {author} {\bibfnamefont {Z.}~\bibnamefont
  {Sefrioui}}, \bibinfo {author} {\bibfnamefont {J.}~\bibnamefont
  {Santamar{\'\i}a}}, \bibinfo {author} {\bibfnamefont {M.}~\bibnamefont
  {Fitzsimmons}}, \bibinfo {author} {\bibfnamefont {S.}~\bibnamefont {Park}}, \
  and\ \bibinfo {author} {\bibfnamefont {M.}~\bibnamefont {Varela}},\
  }\href@noop {} {\bibfield  {journal} {\bibinfo  {journal} {Physical Review
  B}\ }\textbf {\bibinfo {volume} {72}},\ \bibinfo {pages} {140407} (\bibinfo
  {year} {2005})}\BibitemShut {NoStop}%
\bibitem [{\citenamefont {Kavich}\ \emph {et~al.}(2007)\citenamefont {Kavich},
  \citenamefont {Warusawithana}, \citenamefont {Freeland}, \citenamefont
  {Ryan}, \citenamefont {Zhai}, \citenamefont {Kodama},\ and\ \citenamefont
  {Eckstein}}]{kavich2007nanoscale}%
  \BibitemOpen
  \bibfield  {author} {\bibinfo {author} {\bibfnamefont {J.}~\bibnamefont
  {Kavich}}, \bibinfo {author} {\bibfnamefont {M.}~\bibnamefont
  {Warusawithana}}, \bibinfo {author} {\bibfnamefont {J.}~\bibnamefont
  {Freeland}}, \bibinfo {author} {\bibfnamefont {P.}~\bibnamefont {Ryan}},
  \bibinfo {author} {\bibfnamefont {X.}~\bibnamefont {Zhai}}, \bibinfo {author}
  {\bibfnamefont {R.}~\bibnamefont {Kodama}}, \ and\ \bibinfo {author}
  {\bibfnamefont {J.}~\bibnamefont {Eckstein}},\ }\href@noop {} {\bibfield
  {journal} {\bibinfo  {journal} {Physical Review B}\ }\textbf {\bibinfo
  {volume} {76}},\ \bibinfo {pages} {014410} (\bibinfo {year}
  {2007})}\BibitemShut {NoStop}%
\end{thebibliography}%

\end{document}